\documentclass[review,3p,times]{elsarticle}

\usepackage{multirow}
\usepackage{caption}
\usepackage{array}
\usepackage{mdwmath}
\usepackage{mdwtab}
\usepackage{eqparbox}
\usepackage{url}
\usepackage{multirow}

\usepackage{algorithmic}
\usepackage{algorithm}

\usepackage{amssymb}
\usepackage{bm}

\journal{Elsevier}

\usepackage{color}

\begin{document}

\begin{frontmatter}

\title{Uncertainty-Guided Mutual Consistency Learning for Semi-Supervised Medical Image Segmentation}

\author[mymainaddress]{Yichi Zhang}

\author[mymainaddress]{Rushi Jiao}

\author[mymainaddress]{Qingcheng Liao}

\author[mymainaddress]{Dongyang Li}

\author[mymainaddress,mysecondaryaddress,mythirdaddress]{Jicong Zhang\corref{mycorrespondingauthor}}
\cortext[mycorrespondingauthor]{Corresponding author. E-mail address: jicongzhang@buaa.edu.cn }

\address[mymainaddress]{School of Biological Science and Medical Engineering, Beihang University, Beijing, China}
\address[mysecondaryaddress]{Hefei Innovation Research Institute, Beihang University, Hefei, China}
\address[mythirdaddress]{Beijing Advanced Innovation Centre for Biomedical Engineering, Beijing, China}

\begin{abstract}
Medical image segmentation is a fundamental and critical step in many clinical approaches. Semi-supervised learning has been widely applied to medical image segmentation tasks since it alleviates the heavy burden of acquiring expert-examined annotations and takes the advantage of unlabeled data which is much easier to acquire. Although consistency learning has been proven to be an effective approach by enforcing an invariance of predictions under different distributions, existing approaches cannot make full use of region-level shape constraint and boundary-level distance information from unlabeled data. 
In this paper, we propose a novel uncertainty-guided mutual consistency learning framework to effectively exploit unlabeled data by integrating intra-task consistency learning from up-to-date predictions for self-ensembling and cross-task consistency learning from task-level regularization to exploit geometric shape information. The framework is guided by the estimated segmentation uncertainty of models to select out relatively certain predictions for consistency learning, so as to effectively exploit more reliable information from unlabeled data. 
Experiments on two publicly available benchmark datasets showed that: 1) Our proposed method can achieve significant performance improvement by leveraging unlabeled data, with up to 4.13\% and 9.82\% in Dice coefficient compared to supervised baseline on left atrium segmentation and brain tumor segmentation, respectively. 2) Compared with other semi-supervised segmentation methods, our proposed method achieve better segmentation performance under the same backbone network and task settings on both datasets, demonstrating the effectiveness and robustness of our method and potential transferability for other medical image segmentation tasks.
\end{abstract}

\begin{keyword}
Medical Image Segmentation\sep  Semi-Supervised Learning \sep  Uncertainty Estimation \sep  Mutual Consistency Learning
\end{keyword}

\end{frontmatter}

\section{Introduction}

Medical imaging have been widely used in clinical researches, which improves the quality of healthcare by discovering potential lesion and providing diagnostic opinions. Among the various tasks of medical image analysis, medical image segmentation aims to understand images in pixel-level and label each pixel into a certain class, which plays an important role in diagnostic analysis, surgical planning and postoperative analysis, and has attracted the attention of researchers \cite{van2011computer,niessen2016mr,litjens2017survey}.
Based on accurate and robust segmentation results, the morphological attributes of physiological and pathological structures can be quantitatively analyzed, so as to provide useful basis for clinicians to diagnose diseases.
Recently, deep learning-based methods have shown significant improvements and achieved state-of-the-art performances in many medical image segmentation tasks like cardiac segmentation \cite{bernard2018deep,lalande2022deep,xiong2021global}, abdominal segmentation \cite{ma2021abdomenct,heller2020state}, etc.
However, the success of most existing deep learning-based methods relies on a large amount of labeled training data to ease the sub-optimal performance caused by over-fitting and ensure reliable generalization performance on test set, while it is hard and expensive to obtain large-amount well-annotated data in the medical imaging domain where only experts can provide reliable annotations. Besides, many commonly used medical images like computed tomography (CT) and magnetic resonance imaging (MRI) scans are in 3D volumes, which further increase the burden of manual annotation \cite{zhang2022bridging}. 
According to the statistics in \cite{ma2020towards}, it takes about 400$\pm$45 minutes for experts to delineate one CT scan with 250 slices for lung infection segmentation.

To ease the manual labeling burden, significant efforts have been devoted to utilize available annotations efficiently and improve the segmentation performance with low labeling cost \cite{tajbakhsh2020embracing,zhang2021exploiting,cheplygina2019not}. Compared with acquiring expert-examined annotations, unlabeled medical images are relatively easier to obtain. Therefore, implementing medical image segmentation models with only a few labeled images has become an active research topic for clinical applications \cite{semireview}.
In this work, we focus on semi-supervised learning for medical image segmentation by encouraging models to learn from a limited amount of expert-examined labeled data and a large amount of unlabeled data, which is a fundamental, challenging problem and has a high impact on real-world clinical applications.
To utilize unlabeled data, a simple and intuitive method is to assign pseudo annotations to unlabeled data and then train the segmentation model using both labeled and pseudo labeled data. However, model-generated annotations can be noisy and have detrimental effects to the subsequent segmentation model \cite{min2019two}.
Recent impressive progress in semi-supervised medical image segmentation has been focused on incorporating unlabeled data into the training procedure with an unsupervised loss function. Specifically, the mean teacher (MT) model \cite{tarvainen2017mean} has achieved great success by enforcing the consistency of predictions from perturbed inputs between student and teacher models. Following \cite{tarvainen2017mean}, many consistency learning methods \cite{yu2019uncertainty,wang2020double,hang2020local} have been proposed. Besides, another line of researches \cite{li2020shape,luo2021semi,zhang2021dual} focus on building task-level regularization by adding auxiliary task to leverage boundary-based surface mismatch.
However, these consistency-based semi-supervised medical segmentation methods cannot make full use of reliable region-level shape constraint and boundary-level distance information from unlabeled cases.

In this paper, we propose a novel uncertainty-guided mutual consistency learning framework to effectively exploit unlabeled data for semi-supervised medical image segmentation.
We use dual-task backbone network with two output branches to generate segmentation probabilistic maps and signed distance maps simultaneously. For consistency learning, our framework focuses on integrating both the intra-task consistency learning of up-to-date predictions for self-ensembling and cross-task consistency learning from task-level regularization to exploit geometric shape information. 
Besides, our proposed framework is guided by the estimated segmentation uncertainty of models to select out relatively certain predictions for consistency learning, so as to effectively exploit more reliable information from unlabeled data.
We conduct extensive experiments on two publicly available medical image segmentation datasets: Left Atrium Segmentation (LA) and Brain Tumor Segmentation (BraTS).
Experimental results show that our framework can largely improve the segmentation performance by leveraging the unlabeled images and outperform state-of-the-art semi-supervised segmentation methods.

\begin{figure*}[t]
	\includegraphics[width=18cm]{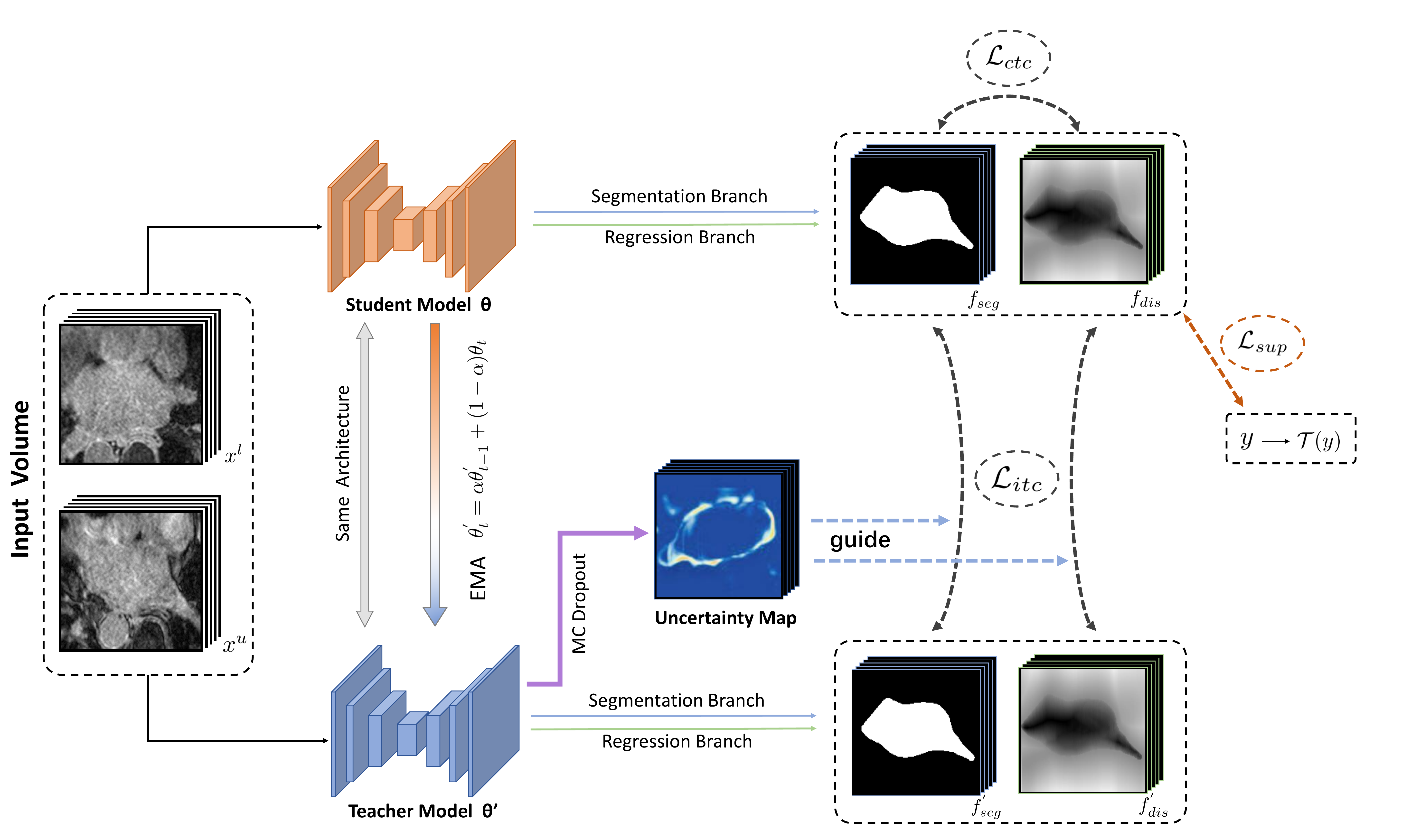}
	\caption{The overview of our proposed uncertainty-guided mutual consistency learning framework for semi-supervised medical image segmentation. The backbone network consists of two different branches for different tasks, where the first branch aims for pixel-wise classification and generates the segmentation probabilistic maps as output, and the second branch aims at level set function regression and regresses the signed distance maps. Following the design of mean teacher framework, the student model is optimized by minimizing the supervised loss on labeled data $\mathcal{D}_{L}$ and cross-task consistency loss on both labeled data $\mathcal{D}_{L}$ and unlabeled data $\mathcal{D}_{U}$. The estimated uncertainty from the teacher model is used to guide the learning of the student model so as to learn more reliable information from unlabeled data for semi-supervised learning.}
	\label{Fig1}
\end{figure*}

\section{Related Work}

\subsection{CNNs for Medical Image Segmentation}

Image segmentation aims to understand the image in pixel level and classify each pixel into a certain class.
Recently, convolutional neural networks (CNNs) have achieved state-of-the-art performance in many medical image segmentation tasks. 
Fully convolutional networks (FCN) \cite{long2015fully} is a landmark in image segmentation. It applied classic convolution neural network to dense prediction and first realized end to end segmentation by replacing fully connected layers with convolutional layers.
Most of widely used methods for medical image segmentation are inspired by U-Net \cite{ronneberger2015u} based on an encoder-decoder structure to extract features at multiple scales. The network architecture fused the features of different scales by concatenating the feature maps of the downsampling layers and the corresponding upsampling layers for subsequent learning. For segmentation of medical volumes, 3D segmentation networks like 3D U-Net \cite{cciccek20163d} and V-Net \cite{milletari2016v} are proposed to use 3D convolution kernels to extract volumetric features.
Besides, many variants of U-Net have been proposed to improve it by designing novel structures and have been applied to many medical image segmentation tasks. \cite{zhou2019unet++,li2018h,zhang2020sau,jin2020raunet,oktay2018attention}.
Isensee \textit{et al.} \cite{isensee2020nnunet} proposed nnU-Net to automatically adapt training strategies and network architectures to a given medical dataset and achieved state-of-the-art performances on many segmentation tasks \cite{antonelli2022medical}, which demonstrated that the basic encoder-decoder structure is still difficult to surpass if the corresponding pipeline is designed adequately.

\subsection{Semi-supervised Medical Image Segmentation}

To reduce the burden of annotation cost, many semi-supervised learning medical image segmentation methods have been proposed by using a limited number of labeled data and an arbitrary amount of unlabeled data.
Existing semi-supervised methods mainly have two categories. The first category is based on pseudo labels, which is an intuitive method by assigning pseudo annotations for unlabeled images, and then using the pseudo labeled images to update the segmentation model.
Bai \textit{et al.} \cite{bai2017semi} iteratively updated the pseudo segmentation labels and network parameters and used conditional random field (CRF) to refine the pseudo labels.
Zhang \textit{et al.} \cite{zhang2017deep} introduced adversarial learning for biomedical image segmentation by encouraging the segmentation output of unlabeled data to be similar to annotations of labeled data.
However, this category ignores the quality of pseudo labels, where model-generated annotations can be noisy and may have detrimental effects to the subsequent segmentation model \cite{min2019two}. 

Another category for semi-supervised learning aims to learn from labeled and unlabeled images simultaneously, with supervised loss for labeled images and unsupervised images for all images. Among these methods, consistency learning is widely used by enforcing an invariance of predictions of input images under different distributions.
For instance, Samuli \textit{et al.} \cite{laine2016temporal} proposed temporal ensembling strategy to use exponential moving average (EMA) predictions for unlabeled data as the consistency targets.
However, maintaining the EMA predictions during the training process is a heavy burden. To issue the problem, Tarvainen \textit{et al.} \cite{tarvainen2017mean} proposed to use a teacher model with the EMA weights of the student model for training. 
Li \textit{et al.} \cite{li2020transformation} applied perturbations like Gaussian noise, randomly rotation and scaling to the input images and encourage the network to be transformation-consistent for unlabeled data.
Yu \textit{et al.} \cite{yu2019uncertainty} extended the mean teacher paradigm with an uncertainty estimation strategy to 
improve the performance of consistency-based model so as to learn from more meaningful and reliable targets during training. 
Luo \textit{et al.} \cite{luo2021efficient} proposed to learn from multi-scale consistency between outputs from different scales for semi-supervised gross target volume segmentation.
Chaitanya \textit{et al.} \cite{chaitanya2020contrastive} proposed novel contrasting strategies to leverage structural similarity and learn distinctive representations of local regions.
Meyer \textit{et al.} \cite{meyer2021uncertainty} proposed an uncertainty-aware temporal self-learning for semi-supervised segmentation of prostate zones and beyond.
Instead of perturbing networks or data for consistency learning, another line of researches focus on building task-level regularization by adding auxiliary task to leverage geometric information with distance maps. 
Li \textit{et al.} \cite{li2020shape} developed a multi-task network to build shape-aware constraints with adversarial regularization.
Luo \textit{et al.} \cite{luo2021semi} combined the regression task with the segmentation task to form a dual-task consistency learning. Zhang \textit{et al.} \cite{zhang2021dual} extended the learning of cross-task consistency with mutual learning of dual-task networks and obtain further performance improvement.

\subsection{Uncertainty Estimation}

For deep neural networks, reliable quantification of uncertainty plays a crucial role in evaluating the confidence of predictions due to the capability to tell when and where the model is likely to make false predictions, especially for medical imaging area. 
Recently, many methods have been developed for uncertainty estimation. Uncertainties for image segmentation are derived from general considerations of the statistical model, from resampling training data sets in ensemble approaches \cite{lakshminarayanan2017simple}, or from modifications like Monte Carlo dropout of the predictive procedure \cite{gal2016dropout}.
For semi-supervised learning, the uncertainty can be used to judge whether the model provides accurate and confident prediction, which can be leveraged to further exploit the unlabeled data and has been applied to many semi-supervised medical image segmentation tasks \cite{yu2019uncertainty,wang2020double}.
Wang \textit{et al.} \cite{wang2021semi} found that Monte Carlo dropout perform better for uncertainty estimation.
In this work, we incorporate the estimated segmentation uncertainty for consistency learning, so as to encourage the model to focus on relatively more certain and reliable information learned from unlabeled data.

\section{Methodology}

An overview of our proposed framework for semi-supervised medical image segmentation is shown in Fig \ref{Fig1}.
To ease the description of the methodology, we formulate the problem of our task as follows.
Given a dataset $\mathcal{D}$ for training, we denote the labeled set with $M$ labeled cases as $\mathcal{D}_{L} = \{x_{i}^{l}, y_{i}\}_{i=1}^{M}$, and the unlabeled set with $N$ unlabeled cases as $\mathcal{D}_{U} = \{x_{i}^{u}\}_{i=1}^{N}$, where $x_{i}^{l}$ and $x_{i}^{u}$ denote the input images and $y_{i}$ denotes the corresponding ground truth of labeled data. 
For semi-supervised segmentation settings, we aim at building a data-efficient deep learning model with the combination of $\mathcal{D}_{L}$ and $\mathcal{D}_{U}$ and making the performance to be comparable to an optimal model trained over fully labeled dataset $\mathcal{D}$.

\subsection{Backbone Network Architecture and Supervised Learning}

\begin{figure}[t]
	\includegraphics[width=16cm]{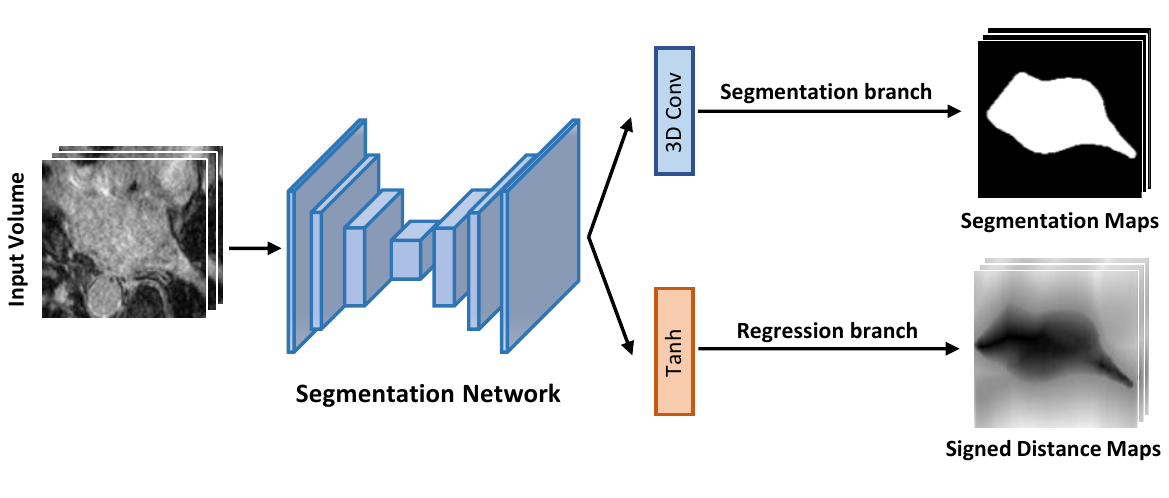}
	\caption{Overview of the backbone network for dual-task learning, where the segmentation branch generates the segmentation probabilistic maps as output and the regression branch generates the signed distance maps as output.}
	\label{SDF}
\end{figure}

In most existing approaches, medical image segmentation can be regarded as a pixel-level classification task to generate segmentation probabilistic maps and assign each pixel into a certain class. 
Other than pixel-wise classification using binary or multi-label masks, another line of researches focus on using signed distance maps by transforming binary masks to gray-level images where the intensities of pixels are changed according to the distance to the closest boundary \cite{ma2020distance}.
As a traditional method to embed object contours in a higher dimensional space, signed distance function (SDF) has recently been incorporated with segmentation CNNs to capture geometric distance information and obtain further improvements \cite{navarro2019shape,dangi2019distance,wang2020deep}.
Specifically, we introduce the transformation from binary ground truth to signed distance maps $\mathcal{T}(x)$ as follows:

\begin{equation}
\mathcal{T}(x)=\left\{
\begin{array}{lllll}
-\inf \limits_{y \in \partial G}\|x-y\|_{2} , & x \in G_{\mathrm{in}}  \\
\\
0, & x \in \partial G \\
\\
+\inf \limits_{y \in \partial G}\|x-y\|_{2} , & x \in G_{\mathrm{out}}  \\
\end{array}\right.
\end{equation}
where $\|x-y\|_{2}$ is the Euclidian distance between voxels $x$ and $y$. $G_{in}$, $\partial G$, $G_{out}$ represent the inside, boundary and outside of the segmentation target, respectively. Generally, SDF takes negative values inside the object and positive values outside the object, while the absolute value of each pixel is defined by the distance to the closest boundary point.
Following the design of classic encoder-decoder architecture like \cite{ronneberger2015u,cciccek20163d,milletari2016v}, 
as shown in Fig \ref{SDF}, an auxiliary regression branch is added to generate the signed distance maps composed by a 3D convolution block followed by a hyperbolic tangent activation, in parallel with the classic segmentation branch to generate the segmentation probabilistic maps.
The task-level difference of two branches can lead to inherent prediction perturbation and encourage the segmentation model to learn different representations of segmentation targets from different perspectives.
For labeled data, supervision from both segmentation branch and regression branch can be utilized for training. Therefore, the supervised segmentation loss can be defined as follows:
\begin{equation}
\mathcal{L}_{sup}(\theta;\mathcal{D}_{L}) = \mathcal{L}_{dice}(f_{seg},y) + \mathcal{L}_{ce}(f_{seg},y) + \mathcal{L}_{dis}(f_{dis},\mathcal{T}(y))
\end{equation}
where $f_{seg}$ and $f_{dis}$ represent the output predictions of segmentation branch and regression branch, respectively. 

\subsection{Intra-Task Consistency Regularization}

For semi-supervised medical image segmentation tasks, the improvement of performance benefits from learning unsupervised knowledge from unlabeled data by generating a supervision signal through unsupervised loss function.
We adopt the mean teacher \cite{tarvainen2017mean} as the main framework of our semi-supervised segmentation approach.
The framework consists of a student model and a teacher model with the same network architecture.
Since ensembling the predictions of the network at different training stages can further bootstrap the quality of learning representations \cite{laine2016temporal}, we update the weights of teacher model $\theta^{'}$ as an exponential moving average (EMA) of the weights of student model $\theta$ as $ \theta^{'}_{t} = \alpha \theta^{'}_{t-1} + (1-\alpha) \theta_{t}$, where $\alpha$ is a hyper-parameter named EMA decay.

For consistency learning, the student model learns from the teacher model by minimizing the combination of supervised segmentation loss and unsupervised consistency loss with respect to the output targets of the teacher model. In our work, we integrate the consistency loss for both segmentation task and regression task.
Following the design in \cite{yu2019uncertainty}, we estimate the uncertainty with Monte Carlo Dropout \cite{kendall2017uncertainties} by performing $T$ stochastic forward passes on the teacher model under random dropout. Therefore, the predictive entropy can be summarized as following equation to approximate the segmentation uncertainty of the model.

\begin{equation}
\hat p_{i} = \frac{1}{T} \sum_{t=1}^{T} p^{t}_{i}  \ \ \ \
u = - \sum_{i=1}^{C} \hat p_{i} log(\hat p_{i})
\end{equation}
where $p^{t}_{i}$ is the prediction logits of class $i$ at the $t^{th}$ time in the forward pass, $C$ is the number of classes in the segmentation tasks,
$\hat p_{i}$ is the average softmax probability of $T$ stochastic passes from teacher model, and $u$ is the estimated segmentation uncertainty.
For a given input, the overall segmentation uncertainty $U$ is the combination of voxel-wise uncertainty $u$.
With the guidance of the estimated uncertainty $U$, we can filter out the relatively unreliable predictions with higher uncertainty and use only the certain predictions for the student model to learn from. 
Therefore, the intra-task consistency loss can be defined as follows:

\begin{equation}
\mathcal{L}_{itc}(\theta;\theta^{'};\mathcal{D})  = \beta \ \frac{ \sum \mathbb I(u<u_{th}) \|f_{seg} - f_{seg}^{'}\|^{2} \  }{ \sum \mathbb I(u<u_{th}) } \\
+ (1 - \beta) \ \frac{\sum \mathbb I(u<u_{th}) \|f_{dis}-f_{dis}^{'}\|^{2}  }{\sum \mathbb I(u<u_{th})} 
\end{equation}
where $(f_{seg}, f_{dis})$ and $(f_{seg}^{'},f_{dis}^{'})$ represent the output of segmentation and regression branch of the student model and the teacher model, respectively. $\mathbb I(\cdot)$ is the indicator function used to select out relatively certain predictions, and $\beta$ is the balance weight of consistency learning between segmentation task and regression task.

\subsection{Cross-Task Consistency Regularization}

Since the task-level difference of two branches can lead to inherent prediction perturbation, different tasks can guide the model to learn different representations of segmentation targets from different perspectives. Under the assumption that predictions of the same input data from different tasks should be consistent when mapped to the same predefined space, in contrast to data-level ensembling consistency, we also regularize cross-task consistency between the outputs of segmentation branch and regression branch to further utilize unlabeled data.
To transform the output of distance maps back to binary segmentation output, we utilize a smooth approximation to the inverse transform as in \cite{luo2021semi,zhang2021dual}, which can be defined by
\begin{equation} 
\mathcal{T}^{-1}(z) =\frac{1}{1+e^{-k \cdot z}} , z \in G_{SDF}
\end{equation}
where $z$ is the value of signed distance maps at voxel $x$, and $k$ is a transform factor selected as large as possible to approximate the transform.
Therefore, the cross-task consistency loss for semi-supervised learning can be defined as follows:

\begin{equation}
\mathcal{L}_{ctc}(\theta;\mathcal{D})  =  \| f_{seg} - \mathcal{T}^{-1}( f_{dis}) \| ^{2}
\end{equation}

\renewcommand{\algorithmicrequire}{ \textbf{Input:}}     
\renewcommand{\algorithmicensure}{ \textbf{Output:}}    
\begin{algorithm}[t]
	\caption{Training procedure of our proposed uncertainty-guided mutual consistency learning framework.}
	\label{Algorithm1}
	\begin{algorithmic}[1]
		\REQUIRE{A batch of $\{x^{l}, y^{l}\}$ from labeled dataset $D_{L}$ and $\{x^{u}\}$ from unlabeled dataset $D_{U}$.}
		\ENSURE{Trained network $\mathcal{N}$ with $\theta_{t}$}
		\STATE $f_{seg}$ and $f_{dis}$ represent the output predictions of segmentation branch and regression branch to generate segmentation probabilistic maps and signed distance maps, respectively.
		\WHILE{not stopping criterion}
		\STATE ($x_{i}^{l}, y_{i}$), ($x_{i}^{u}$) $\leftarrow$ sampled from $D_{L}$ and $D_{U}$
		\STATE Generate output segmentation maps $f_{seg}$, output distance maps $f_{dis}$ and estimated uncertainty $U$
		\STATE Calculate supervised segmentation loss $\mathcal{L}_{sup}$ as Eq. (2)
		\STATE Calculate intra-task consistency losses $\mathcal{L}_{itc}$ as Eq. (4) 
		\STATE Calculate cross-task consistency losses $\mathcal{L}_{ctc}$ as Eq. (6) 
		\STATE Update the student model's weights $\theta$ with \\
		$\mathcal{L} = \mathcal{L}_{sup} + \lambda_{i} \mathcal{L}_{itc} + \lambda_{c} \mathcal{L}_{ctc}$
		\STATE Update the teacher model's weights $\theta^{'}$ with exponential moving average (EMA) of the student model’s weights as $ \theta^{'}_{t} = \alpha \theta^{'}_{t-1} + (1-\alpha) \theta_{t}$   
		\ENDWHILE 
	\end{algorithmic}
\end{algorithm}

\subsection{Overall Training Procedure}

The overall training objective of our proposed framework is to minimize the weighted sum of supervised segmentation loss $\mathcal{L}_{sup}$, intra-task consistency loss $\mathcal{L}_{itc}$ and cross-task consistency loss $\mathcal{L}_{ctc}$.
The student model first explicitly learns from labeled data $D_{L}$ via the supervised segmentation loss $\mathcal{L}_{sup}$.
Meanwhile, the student model also acquires useful information from unlabeled data $D_{L}$ with the guidance of estimated uncertainty.
Therefore, the task can be formulated as training the network by minimizing the following functions.
\begin{equation}
\min \limits_{\theta} \mathcal{L}_{sup}(\theta;\mathcal{D}_{L}) + \lambda_{i} \mathcal{L}_{itc}(\theta;\theta^{'};\mathcal{D}) + \lambda_{c} \mathcal{L}_{ctc}(\theta;\mathcal{D})
\end{equation}
where $\lambda_{i}$ and $\lambda_{c}$ are ramp-up weighting coefficients that control the trade-off between supervised and unsupervised loss, so as to mitigate the disturbance of consistency loss at early training stage.
The training objective of our proposed uncertainty-guided mutual consistency learning framework can be formulated as Algorithm \ref{Algorithm1}.

\section{Experiments}

\subsection{Datasets and Experimental Setup}

We extensively evaluate our proposed method on two public datasets. 
The first dataset is Left Atrium (LA) dataset from Atrial Segmentation Challenge\renewcommand{\thefootnote}{1}\footnote{http://atriaseg2018.cardiacatlas.org/data/} \cite{xiong2021global}. 
The dataset contains 100 3D gadolinium-enhanced MR imaging scans (GE-MRIs) and corresponding segmentation masks of left atrium for training and validation. These scans have an isotropic resolution of $0.625 \times 0.625 \times 0.625$ $mm^{3}$. Following the same task setting in \cite{yu2019uncertainty}, we split the 100 scans into 80 scans for training and 20 scans for testing, and apply the same pre-processing methods. Out of the 80 training scans, we use the same 20\%/16 scans as labeled data and the remaining 80\%/64 scans as unlabeled data for semi-supervised segmentation task. 
The second dataset is Brain Tumor Segmentation (BraTS) 2019 dataset \renewcommand{\thefootnote}{2}\footnote{https://ieee-dataport.org/competitions/brats-miccai-brain-tumor-dataset} \cite{hdtd-5j88-20}. The dataset contains multi-institutional preoperative MRI of 335 glioma patients, where each patient has four modalities of MRI scans including T1, T1Gd, T2 and T2-FLAIR with neuroradiologist-examined labels. We use T2-FLAIR for whole tumor segmentation since such modality can better manifest the malignant tumors \cite{zeineldin2020deepseg}. 
All the scans are resampled to the same resolution of $1 \times 1 \times 1$ $mm^{3}$ with intensity normalized to zero mean and unit variance.
In our experiments, we split the dataset into 250 scans for training, 25 scans for validation and the remaining 60 scans for testing. Among the 250 training scans, we conduct experiments under two different settings with 10\%/25 and 20\%/50 scans as labeled data and the remaining scans as unlabeled data.

\begin{table*}[]
	\caption{Ablation analysis of our mutual consistency learning framework on LA dataset. The arrows of evaluation metrics indicate which direction is better.} \label{Table_ab}
	\centering
	\renewcommand\arraystretch{1}
	\begin{tabular}{c|cc|cc|c|c|c|c}
		\hline 	\hline
		\multirow{2}{*}{Method} &\multicolumn{2}{c|}{Supervised Loss} & \multicolumn{2}{c|}{Consistency Loss} &  \multicolumn{4}{c}{Metrics}  \\ 
		\cline{2-9} & $\mathcal{L}_{seg}$    &   $\mathcal{L}_{dis}$  &   $\mathcal{L}_{itc}$    &   $\mathcal{L}_{ctc}$      
		&Dice[\%] $\uparrow$	&Jaccard[\%] $\uparrow$		&ASD[voxel] $\downarrow$	&95HD[voxel]  $\downarrow$  \\ \hline
		(1) & $\checkmark$  & - &  - & -      & 86.03     & 76.06        & 3.51        & 14.26  \\
		(2) & $\checkmark$  & $\checkmark$ &  - & -      & 87.88     & 78.77  & 2.81  & 10.25 \\ \hline
		(3) & $\checkmark$  & - &  $\checkmark$ & - &   88.68     & 79.90       & 2.71       & 9.07    \\ 
		(4) & $\checkmark$  & $\checkmark$ &  $\checkmark$ & -  & 89.15 & 80.58 & 2.03 & 8.15    \\
		(5) & $\checkmark$  & $\checkmark$ &  -  &  $\checkmark$      & 89.42     & 80.89        & 2.10        & 7.32    \\
		(6) & $\checkmark$  & $\checkmark$ &  $\checkmark$ & $\checkmark$  & \textbf{90.16} & \textbf{82.18} & \textbf{1.98} & \textbf{6.50} \\ \hline \hline
	\end{tabular}
\end{table*}

\subsection{Implementing Details and Evaluation Metrics}

All of our experiments are implemented in Python with PyTorch, using an NVIDIA Tesla V100 GPU with 32GB memory.
We adopt the same V-Net \cite{milletari2016v} as the backbone structure for all experiments to ensure a fair comparison. 
To control the balance between supervised segmentation loss and unsupervised consistency loss, we use a Gaussian ramp-up function $ \lambda(t)=0.1*e^{-5(1-t/t_{max})}$ as \cite{tarvainen2017mean} in our experiments, where $t$ represents the current number of iterations and $t_{max}$ represents the maximum number of iterations.
We use the Stochastic Gradient Descent (SGD) optimizer to update the network parameters with an initial learning rate of 0.01 decayed by 0.1 every 2500 iterations. The maximum training iterations is set to 6,000. The batch size is set to 4, consisting of 2 labeled images and 2 unlabeled images in each mini-batch. We randomly crop $112 \times 112 \times 80$ sub-volumes as the network input and the final segmentation results are obtained using a sliding window strategy.
We use the standard data augmentation techniques on-the-fly to avoid overfitting during the training procedure \cite{yu2017automatic}, including randomly flipping, and rotating with 90, 180 and 270 degrees along the axial plane.
To quantitatively evaluate the segmentation results, we use four complementary evaluation metrics. Dice similarity coefficient (Dice) and Jaccard Index (Jaccard), two region-based metrics, are used to measure the region mismatch. Average surface distance (ASD) and 95\% Hausdorff Distance (95HD), two boundary-based metrics, are used to evaluate the boundary errors between the segmentation results and the ground truth.

\begin{table}[]
	\caption{Experimental results of different balance weight $\beta$ of segmentation task and regression task for self-ensembling on LA dataset. } \label{Table_beta}
	\centering
	\renewcommand\arraystretch{0.9}
	\begin{tabular}{c|c|c|c|c}
		\hline 	\hline
		\multirow{2}{*}{$\beta$}  &  \multicolumn{4}{c}{Metrics}  \\ 
		\cline{2-5} &Dice[\%] $\uparrow$	&Jaccard[\%] $\uparrow$		&ASD[voxel] $\downarrow$	&95HD[voxel]  $\downarrow$ \\ \hline	
		0    & 89.60          & 81.30          & 2.51          & 8.66 \\
		0.25 & 89.86          & 81.75          & 2.22          & 7.18 \\
		0.5  & 90.07          & 82.03          & 2.17          & 7.57 \\
		0.75 & \bm{$90.16$}   & \bm{$82.18$}   & \bm{$1.98$}   & \bm{$6.50$} \\
		1    & 89.99          & 81.93          & 2.74          & 7.95 \\    \hline \hline
	\end{tabular}
\end{table}

\begin{figure*}[t]
	\centering
	\includegraphics[width=14cm]{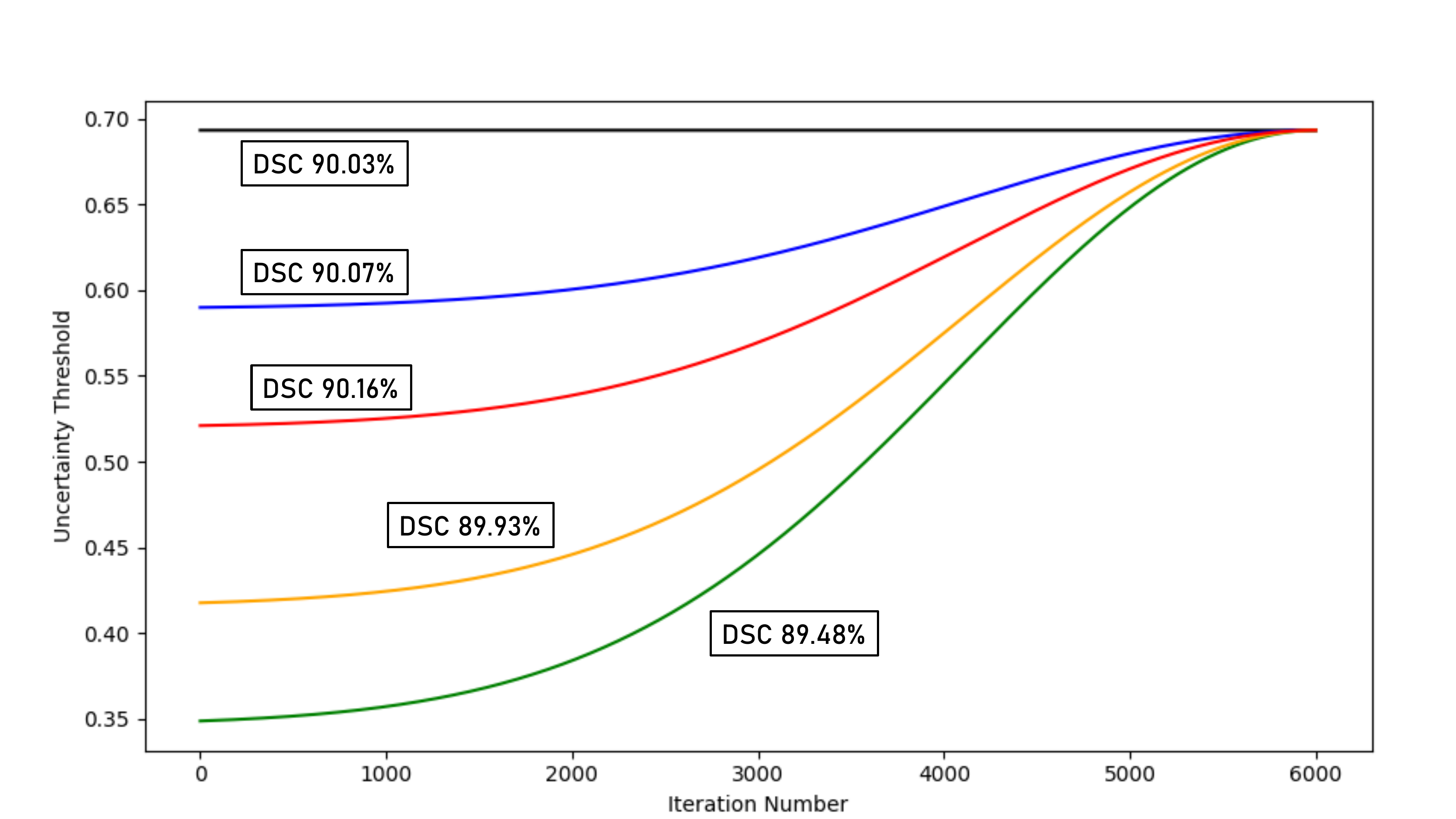}
	\caption{Comparison of Dice performance with different uncertainty threshold $u_{th}$ on left atrium segmentation dataset.}
	\label{uth}
\end{figure*}

\begin{figure*}[t]
	\centering
	\includegraphics[width=16cm]{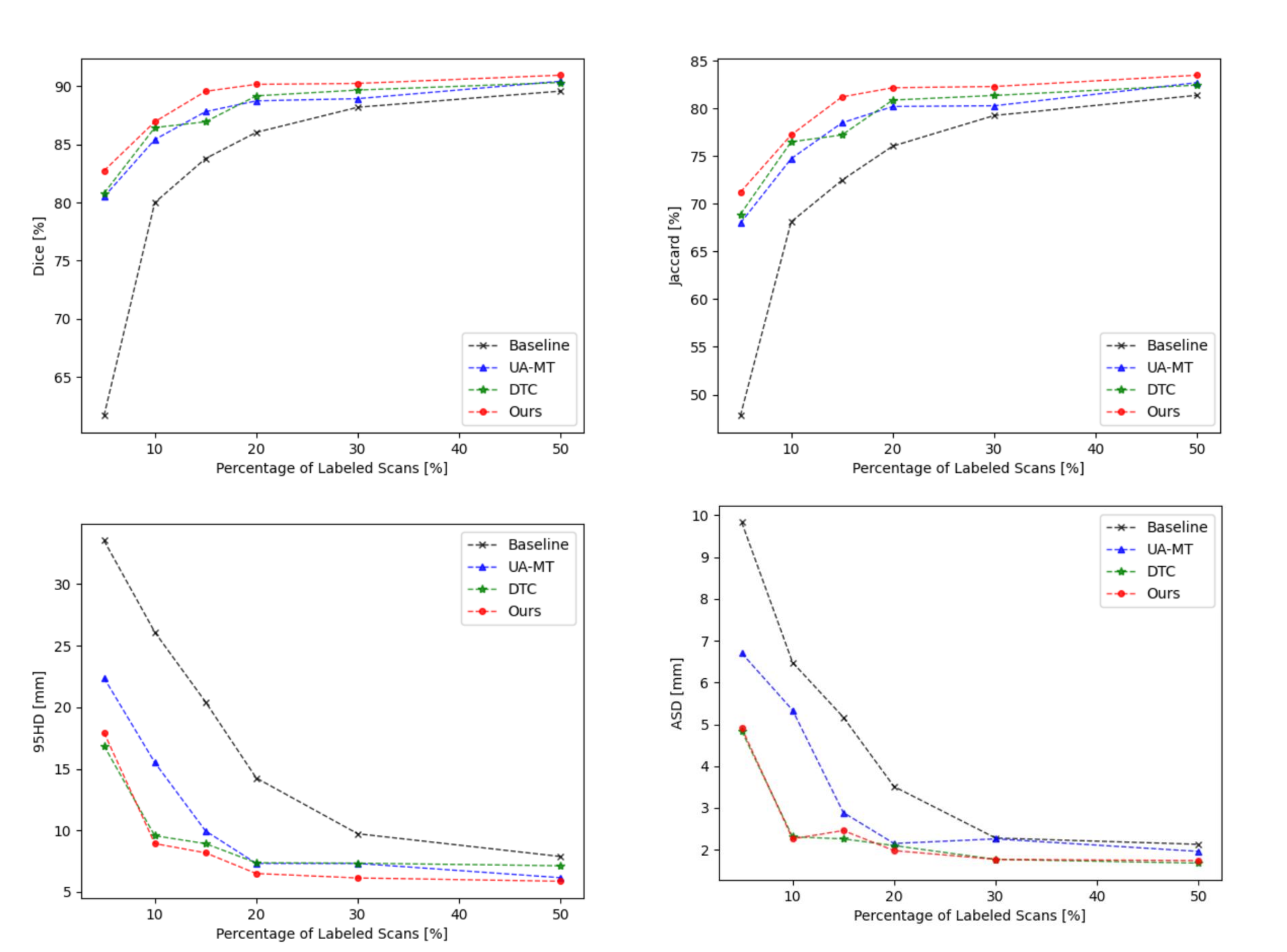}
	\caption{Comparison of segmentation performance of our proposed method (in red) with comparison to fully supervised baseline (in black), UA-MT (in blue) and DTC (in green) using different percentages of labeled data (5\%, 10\%, 15\%, 20\%, 30\% and 50\%) on LA dataset.}
	\label{Percentage}
\end{figure*}

\begin{table*}[!t]
	\caption{Quantitative comparison between our methods and other semi-supervised methods on LA dataset. All the models use the same V-Net as the backbone. The first and second rows are upper-bound performance and fully supervised baseline. * and ** denote the p-value < 0.05 and p-value < 0.01 based on paired t-test when comparing the proposed method with others.} \label{Table_sota_la}
	\centering
	\renewcommand\arraystretch{1}
	\begin{tabular}{c|c|c|c|c|c|c}
		\hline 	\hline
		\multirow{2}{*}{\bf{Method}} & \multicolumn{2}{c|}{\textbf{Scans used}} & \multicolumn{4}{c}{\bf{Metrics}}\\
		\cline{2-7}		&Labeled 	&Unlabeled	&Dice[\%] $\uparrow$ &Jaccard[\%] $\uparrow$	&ASD[voxel] $\downarrow$	&95HD[voxel]  $\downarrow$  \\ \hline
		Supervised baseline     & 16            & 0              & $86.03^{**}$     & $76.06^{**}$        & $3.51^{**}$        & $14.26^{**}$        \\ 
		TCSE \cite{li2020transformation} & 16            & 64             & $88.15^{**}$     & $79.20^{**}$        & $2.44^{*}$        & $9.57^{*}$         \\
		MT \cite{tarvainen2017mean}      & 16            & 64             & $88.23^{**}$     & $79.29^{**}$       & $2.73^{*}$        & $10.64^{**}$         \\
		UA-MT  \cite{yu2019uncertainty}   & 16            & 64             & $88.88^{*}$     & $80.21^{*} $       & $2.26$        & $7.32$         \\
		Entropy Mini \cite{vu2019advent} & 16 & 64 &  $88.45^{*}$ & $79.51^{**}$  & $3.72 ^{**}$ & $14.14^{**}$  \\
		SASS  \cite{li2020shape}          & 16            & 64             & $89.54$     & $81.24 $       & $ 2.20$       & $8.24^{*}$         \\
		DUWM  \cite{wang2020double}  & 16            & 64             & $ 89.65 $    & $ 81.35 $       & $ 2.03 $       & $7.04$         \\
		URPC  \cite{luo2021efficient}  &  16	  &  64	  & $88.74^{*}$  &	$79.93^{*}$  &	$3.66^{**}$  &	$12.73^{**}$ \\
		DTC \cite{luo2021semi}           & 16            & 64             & $89.42^{*}$     & $80.89^{*}$        & $2.10$        & $7.32$         \\
		\textbf{UG-MCL (Ours)} & 16            & 64        & \bm{$90.16$} & \bm{$82.18$} & \bm{$1.98$} & \bm{$6.50$}    \\ \hline
		Supervised upper-bound   & 80            & 0              & $91.14$     & 83.82        & 1.52        & 5.75         \\ \hline \hline
	\end{tabular}
\end{table*}

\begin{figure*}[t]
	\centering
	\includegraphics[width=16cm]{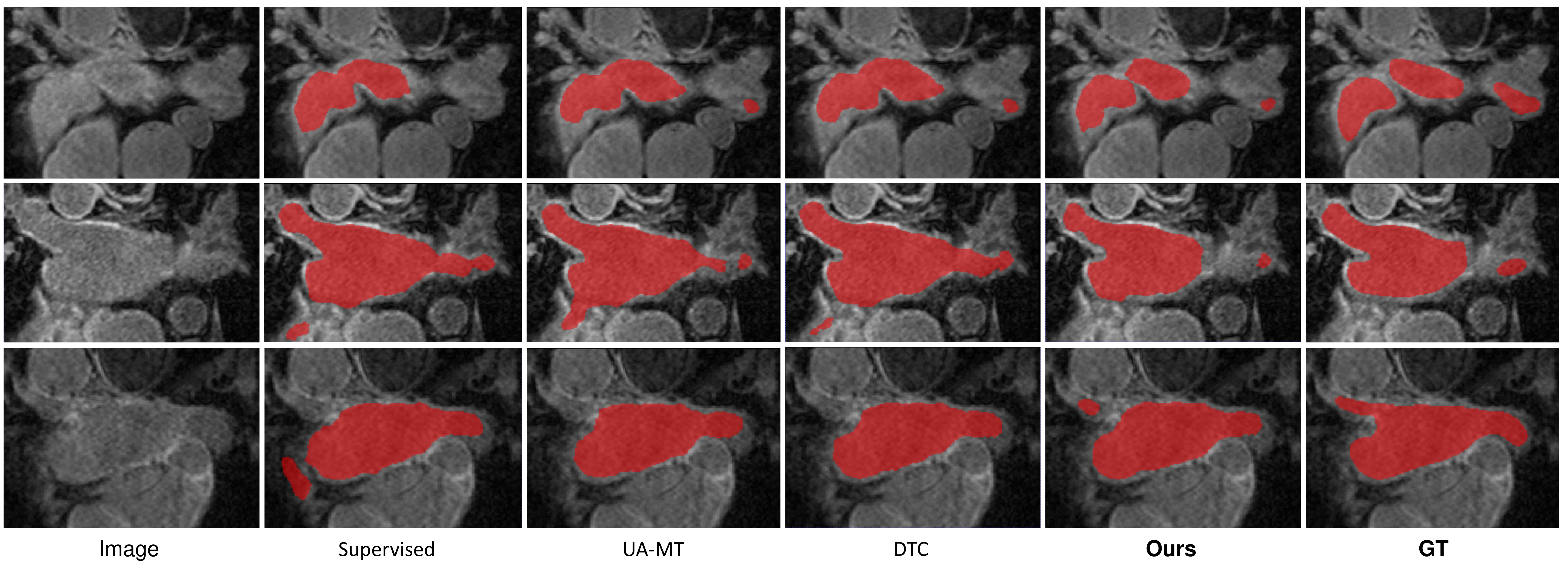}
	\caption{Visual comparison of the left atrium segmentation results of our proposed method with comparison to fully supervised baseline and other state-of-the-art semi-supervised methods using 20\% labeled data.}
	\label{LA_results}
\end{figure*}

\subsection{Ablation Experiments}

Since the classic consistency learning procedure of self-ensembling is based on the segmentation predictions, for our segmentation network with dual-task outputs , we use a balance weight $\beta$ to control the consistency learning between segmentation task and regression task. We conduct experiments to evaluate the selection of $\beta$ in our mutual consistency learning framework.
The quantitative results are shown in Table \ref{Table_beta}. When $\beta$ = 1 or 0, the model achieves lower performance because the ensembling of predictions is only based on segmentation or regression branch, while another branch is neglected. It can be found that when $\beta$ = 0.75, the framework can achieve the best performance. Therefore, we use $\beta$ = 0.75 in our framework in the following experiments.
Besides, following the selection of uncertainty threshold in \cite{yu2019uncertainty} with Gaussian ramp-up paradigm, we conduct experiments to show the influence of different choices for uncertainty threshold. The uncertainty threshold selection with corresponding Dice score is illustrated in Figure \ref{uth}. Specifically, the upper one in black with uncertainty threshold fixed to $U_{max}$ for all iterations. Based on the results, we use uncertainty threshold in red ramping from $\frac{3}{4}U_{max}$ to $U_{max}$ in our framework.

To evaluate the effectiveness of our proposed framework, we conduct ablation studies by removing different components in our method. All the experiments are performed on LA dataset using 16 labeled scans and 64 labeled scans for comparison. 
The testing results of ablation experiments are shown in Table \ref{Table_ab}. Experiment (1) and (2) are the supervised baseline of V-Net trained with only labeled data. We can observe that by utilizing unlabeled data, all semi-supervised methods can significantly improve the segmentation performance compared with supervised baseline results. In experiment (3), we only activate the classic segmentation branch for consistency learning, while the regression branch is removed. In experiment (4), the supervision based on the transformed distance maps of ground truth is utilized to regularize the training. We can observe that adding supervision on distance maps further improves the segmentation by 0.47\% in Dice and 0.68\% in Jaccard. In experiment (5), when removing the intra-task consistency learning of teacher-student framework, the model degenerates into a dual-task V-Net with only cross-task consistency learning.
From the results, we can observe that with the integration of intra-task consistency and cross-task consistency for mutual consistency learning, the performance of our framework is further promoted.
Besides, we also conduct paired t-test to validate the significance of the improvements between our method and the ablation studies. The results show that all the improvements are statistically significant at $p < 0.05$, demonstrating the effectiveness of our framework.

\subsection{Performance of Using Different Percentages of Labeled Data}

We performed a study on data utilization efficiency of our semi-supervised framework with comparison to fully supervised baseline and other two semi-supervised methods including uncertainty-aware mean teacher (UA-MT) \cite{yu2019uncertainty} and dual task consistency learning (DTC) \cite{luo2021semi}.
We conduct experiments of using different percentages of labeled data of LA dataset. The visualization of segmentation results are presented in Figure \ref{Percentage}.
It can be observed that all semi-supervised methods consistently perform better than the supervised baseline in different labeled data settings. Besides, our method outperforms other semi-supervised segmentation methods consistently with different percentages of labeled data, demonstrating the superiority of our proposed framework. Compared with the supervised baseline, our proposed method obtains improvement of 21.07\%, 6.95\%, 5.78\%, 4.13\%, 2.05\% and 1.36\% in Dice by using 5\%, 10\%, 15\%, 20\%, 30\% and 50\% of labeled training data, respectively.
We notice that the performance improvement of semi-supervised learning gradually narrows with the increase of labeled data, which is in line with the common sense.

\subsection{Comparison Experiments with Other Semi-Supervised Segmentation Methods on LA Dataset}

To demonstrate the effectiveness of our method, a comprehensive comparison with existing methods is conducted on LA dataset.
We evaluate our method with comparisons to several recent state-of-the-art semi-supervised segmentation methods, including transformation-consistent self-ensembline (TCSE) \cite{li2020transformation}, mean teacher (MT) \cite{tarvainen2017mean}, uncertainty-aware mean teacher (UA-MT) \cite{yu2019uncertainty}, entropy minimization (Entropy Mini) \cite{vu2019advent}, shape-aware semi-supervised segmentation (SASS) \cite{li2020shape}, double uncertainty weighted method (DUWM) \cite{wang2020double}, dual task consistency learning (DTC) \cite{luo2021semi} and uncertainty rectified pyramid consistency (URPC) \cite{luo2021efficient}.
To ensure a fair comparison, we used the same network backbone in these methods. 
Fig \ref{LA_results} presents some visualization of segmentation results from different semi-supervised segmentation methods. It can be observed that our proposed method generates more accurate and complete segmentation than other methods.
Table \ref{Table_sota_la} shows the quantitative results on the LA dataset. As a contrast, we also conduce experiments of V-Net under fully-supervised settings with 20\% and all labeled data as the lower-bound and upper-bound references for the task. Compared with semi-supervised learning settings, only labeled scans are used for the lower-bound subtask, while both labeled and unlabeled scans with annotations are used for the upper-bound subtask. 
It can be observed that by efficiently leveraging unlabeled data for training, our proposed method can achieve significant performance gains from 86.03\% to 90.16\% of Dice and 76.06\% to 82.18\% of Jaccard. Meanwhile, our method obtains comparable results of 90.16\% in Dice compared with the upper-bound performance of 91.14\%, and significantly outperforms other semi-supervised segmentation methods in terms of all the four evaluation metrics.

\begin{table*}[!t]
	\caption{Quantitative comparison between our methods and other semi-supervised methods on BraTS 2019 dataset. All the models use the same V-Net as the backbone. The first and second rows are upper-bound performance and fully supervised baseline. * and ** denote the p-value < 0.05 and p-value < 0.01 based on paired t-test when comparing the proposed method with others.} \label{Table_sota_brats}
	\centering
	\renewcommand\arraystretch{1}
	\begin{tabular}{c|c|c|c|c|c|c}
		\hline 	\hline
		\multirow{2}{*}{\bf{Method}} & \multicolumn{2}{c|}{\textbf{Scans used}} & \multicolumn{4}{c}{\bf{Metrics}}\\
		\cline{2-7}		&Labeled 	&Unlabeled	&Dice[\%] $\uparrow$ &Jaccard[\%] $\uparrow$	&ASD[voxel] $\downarrow$	&95HD[voxel]  $\downarrow$  \\ \hline
		Supervised baseline    & 25            & 0             & $73.00^{**}$     & $59.96^{**}$        & $2.96^{**}$        & $42.64^{**}$  \\
		MT \cite{tarvainen2017mean}      & 25            & 225             & $81.21^{**}$     & $70.83^{**}$      & $2.45$        & $14.72^{**}$     \\
		UA-MT  \cite{yu2019uncertainty}   & 25            & 225             & $80.85^{**}$     & $70.32 ^{**}$   & $2.57^{*}$        & $14.61^{*}$    \\
		Entropy Mini \cite{vu2019advent} & 25 & 225 & $81.74^{*}$ & $71.42^{*}$ & $2.37$ & $13.71^{*}$ \\
		URPC  \cite{luo2021efficient}  &  25	  &  225	  & $81.80^{*}$  &	$71.63$  &	$2.48$ &	$11.50$   \\
		DTC \cite{luo2021semi}           & 25            & 225      & $81.96$ & $71.84$ & $2.43$    & $12.08$    \\
		\textbf{UG-MCL (Ours)} & 25            & 225      & \bm{$82.82$}  & \bm{$72.77$}  &  \bm{$2.30$}  & \bm{$11.29$}   \\  \hline 
		
		Supervised baseline    & 50            & 0             & $76.14^{**}$ & $64.15^{**}$ & $2.70^{**}$ & $36.01^{**}$  \\
		MT \cite{tarvainen2017mean}      & 50            & 200       & $82.38^{*}$ & $72.32^{**}$  & $2.21$ & $14.54$     \\
		UA-MT  \cite{yu2019uncertainty}   & 50            & 200     & $81.57^{**}$ & $71.42^{**}$ & $2.49^{*}$ & $13.98$               \\
		Entropy Mini \cite{vu2019advent} & 50            & 200  & $82.37^{*}$  & $72.28^{*}$ & $2.30$ &  $15.83^{*}$ \\
		URPC  \cite{luo2021efficient} & 50            & 200	  & $82.80$ & $72.72^{*}$ & $2.72^{**}$ & $12.48$  \\
		DTC \cite{luo2021semi}           & 50            & 200           & $82.78^{*}$ & $72.47^{**}$ & \bm{$2.20$} & $13.43$         \\
		\textbf{UG-MCL (Ours)} & 50            & 200    & \bm{$83.61$}  & \bm{$73.98$}  &  $2.26$ & \bm{$11.44$} \\  \hline \hline
	\end{tabular}
\end{table*}

\begin{figure*}[t]
	\centering
	\includegraphics[width=16cm]{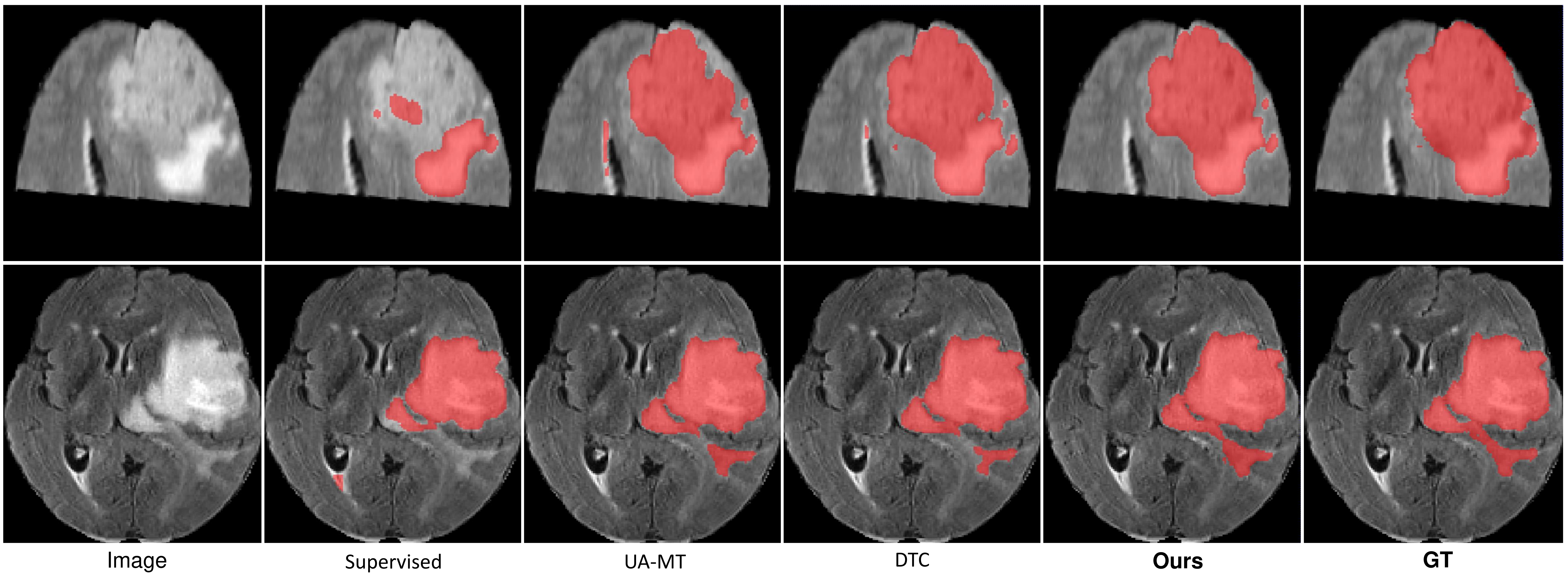}
	\caption{Visual comparison of the brain tumor segmentation results of our proposed method with comparison to fully supervised baseline and other state-of-the-art semi-supervised methods using 10\% labeled data.}
	\label{BraTS_results}
\end{figure*}

\subsection{Comparison Experiments with Other Semi-Supervised Segmentation Methods on BraTS Dataset}

To further validate the effectiveness and generalization ability of our proposed method, we conduct experiments on BraTS 2019 dataset with comparison to existing methods under 10\% and 20\% labeled data settings. Table \ref{Table_sota_brats} presents the segmentation results of our proposed method and other semi-supervised segmentation methods on the dataset. We can observe that comparing with left atrium segmentation, the brain tumor segmentation is a more challenging task due to the irregular shape and boundary of tumors. With 10\% and 20\% labeled data for training, the supervised baseline only yields an average Dice of 73.00\% and 76.14\%, respectively.
For semi-supervised segmentation settings, although the performances are comparable among all the comparing methods, our proposed method outperforms other state-of-the-art semi-supervised methods with different amounts of labeled data, demonstrating the superiority of our framework to effectively exploit the information from unlabeled data. In Fig \ref{BraTS_results}, we present some of the brain tumor segmentation results using 10\% labeled data. We can observe that our proposed method generates more accurate predictions compared with other methods, which further demonstrates the effectiveness of our proposed method.

\section{Conclusion}

Despite existing deep learning-based medical image segmentation methods have achieved great success, it is also limited by requiring large amount of expert-examined annotations. Semi-supervised segmentation by encouraging segmentation models to utilize unlabeled data which is much easier to acquire havs shown the potential to deal with this challenge \cite{semireview}.
In this paper, we present a novel uncertainty-guided mutual consistency learning framework for semi-supervised medical image segmentation.
We use dual-task backbone network with two output branches to generate segmentation probabilistic maps and signed distance maps simultaneously.
To effectively exploit unlabeled data for training, our framework integrates intra-task consistency learning from up-to-date predictions for self-ensembling and cross-task consistency learning from task-level regularization to exploit geometric shape information. 
Our proposed framework is guided by the estimated segmentation uncertainty of models to select out relatively certain predictions for consistency learning, so as to effectively exploit more reliable information from unlabeled data.
Extensive experiments on two public medical image segmentation datasets demonstrate the superiority of our proposed method over other semi-supervised learning methods. Based on the experimental results, our proposed method can achieve significant performance improvement by leveraging unlabeled data, with up to 4.13\% and 9.82\% in Dice coefficient compared to supervised baseline on left atrium segmentation and brain tumor segmentation, respectively. Besides, our proposed method achieve better segmentation performance compared with other semi-supervised segmentation methods under the same backbone network and task settings on both datasets.
The proposed method has the potential to be applied to further clinical applications to ease the burden of annotation cost.
In this work, one limitation is we only focus on single-class segmentation tasks on relatively small datasets. In the future work, we aim to focus on more challenging multi-class segmentation tasks on more diverse datasets like \cite{ma2021abdomenct} to further improve and validate the performance of semi-supervised medical image segmentation methods in real-world clinical applications.

\section*{Declaration of Competing Interest}
There are no conflicts of interest.

\section*{Acknowledgement}
This work is supported in part by the National Key Research and Development Program of China (2016YFF0201002), and in part by the University Synergy Innovation Program of Anhui Province (GXXT-2019-044).

\bibliography{reference}

\end{document}